\documentclass[11pt,twocolumn,a4paper]{article}
\usepackage[utf8]{inputenc}
\usepackage{epsf,exscale,times}
\usepackage[english]{babel}
\usepackage{graphicx}
\usepackage{amsmath}
\usepackage{fancyhdr}
\usepackage{cleveref}
\usepackage{amssymb}
\usepackage{epsf}
\usepackage{tikz}
\usepackage[dvips]{epsfig}
\usepackage{lscape,multicol}
\usepackage{latexsym}
\usepackage{rotating}
\usepackage{subcaption}
\usepackage{authblk}

\def\comp {{\mathchoice {\setbox0=\hbox{$\displaystyle\rm C$}\hbox{\hbox
				to0pt{\kern0.4\wd0\vrule height0.96\ht0\hss}\box0}}
		{\setbox0=\hbox{$\textstyle\rm C$}\hbox{\hbox
				to0pt{\kern0.4\wd0\vrule height0.96\ht0\hss}\box0}}
		{\setbox0=\hbox{$\scriptstyle\rm C$}\hbox{\hbox
				to0pt{\kern0.4\wd0\vrule height0.96\ht0\hss}\box0}}
		{\setbox0=\hbox{$\scriptscriptstyle\rm C$}\hbox{\hbox
				to0pt{\kern0.4\wd0\vrule height0.96\ht0\hss}\box0}}}}
\def\zent {{\mathchoice {\hbox{$\sf\textstyle Z\kern-0.4em Z$}}
		{\hbox{$\sf\textstyle Z\kern-0.4em Z$}}
		{\hbox{$\sf\scriptstyle Z\kern-0.3em Z$}}
		{\hbox{$\sf\scriptscriptstyle Z\kern-0.2em Z$}}}}

\crefname{figure}{fig.}{fig.}

\newcommand\blfootnote[1]{%
  \begingroup
  \renewcommand\thefootnote{}\footnote{#1}%
  \addtocounter{footnote}{-1}%
  \endgroup
}

\title{Multi-criteria Analysis for Evaluation of Adaptive Radar Resource Management algorithms on a Naval setting with and without clutter}
\date{}

\author[1]{Christophe Labreuche}
\author[1]{Cédric Buron}
\author[2]{Peter Moo}
\author[3]{Frédéric Barbaresco}
\affil[1]{Thales Research \& Technology\\
    1, avenue Augustin Fresnel, 91767 Palaiseau cedex, France\\
		email: \{christophe.labreuche,cedric.buron\}@thalesgroup.com}
\affil[2]{Radar Systems Section, Defence Research and Development Canada\\
Ottawa, Ontario, Canada\\
    		email: PETER.MOO@forces.gc.ca}
\affil[3]{Thales Land \& Air Systems, Surface Radar Business Line, \\  Hameau de Roussigny, F-91470 Limours, France  \\
        email: frederic.barbaresco@thalesgroup.com}

\begin{document}
	
	\maketitle
	

	\begin{abstract}\blfootnote{C. Labreuche, C. L. R. Buron, P. Moo and F. Barbaresco, ``Multi-criteria Analysis for Evaluation of Adaptive Radar Resource Management algorithms on a Naval setting with and without clutter,'' \textit{2019 International Radar Conference (RADAR)}TOULON, France, 2019, pp. 1--6, doi: 10.1109/RADAR41533.2019.171240.}Multifunction radars (MFR) are met with complex capability requirements, involving various kinds of targets and saturating scenarios. In order to achieve these goals, radar systems use Active Electronically Scanned Array (AESA) to switch between their functions. This complexity makes it difficult to assess their performance both taking into account their functional requirements and the induces cost. In this paper, we first introduce Figures of Merit for phased array radar operation and performance. We compute these metrics on a set of test scenarios of varying complexity, involving different kinds of targets, and we aggregate them into a global notation of MFR radar performances. To achieve our purpose, we rely on the domain of Multi Criteria Decision Analysis (MCDA), and more precisely 2-additive Choquet integral.
	\end{abstract}
	\section{Introduction}
	\label{Sintro}
	Naval multi-function radars need to conduct multiple functions, among which surveillance, tracking, fire control and various other functions \cite{Briheche2018,Briheche2018a,Pilte}. In order to achieve its goal, the radar relies on a Radar Resource Management (RRM) technique. Its purpose is to allocate the resources of the radar in an efficient way, depending on the radar's mission (noncooperative target recognition, kill assessment\dots). Traditionally for naval radars, the control and the scheduling of the task do not vary over time, \textsl{i.e.} the RRM algorithm is non-adaptive. However, nowadays increasingly complex situation may require the use of adaptive RRM algorithms \cite{mooding15,baretal03A}, which changes the resource management depending both on the target and the interference environment \cite{jeanbar14}. In order to evaluate the performance of these approaches in comparison to non-adaptive algorithms with regards to complexity and robustness, further study is required \cite{barlab09}.

	
	The contribution of this paper is twofold: we first refine the general Multi-Criteria Decision Analysis (MCDA) approach for assessing the results of RRM simulations, and we then apply this approach to a set of naval scenarios.
	
	
	In order to achieve our purpose, we evaluate several candidate RRM algorithms. In order to take all the functions of the radar into account, the RRM algorithm shall be measured on several metrics. The metrics consist of the track completeness of the different kinds of targets. Multi Criteria Decision Analysis is used to define the overall utility of a RRM algorithm when compared to another one. Suppose we make measurements on 2 radars using two metrics, and the first algorithm is better than the second one on the first one, while the second algorithm is better on the second metric. In this case, there is no trivial way to choose which of the algorithm is the best one. MCDA is used in that kind of situations to solve this problem, while relying on the knowledge of a Subject Matter Expert (SME). Such appraoach has already been proposed by the authors in \cite{barlab09,lab17v} with the support of a tool called MYRIAD \cite{lab05v}.
	
\begin{figure}
        \centering
        \begin{subfigure}[b]{.45\linewidth}
            \includegraphics[width=\textwidth]{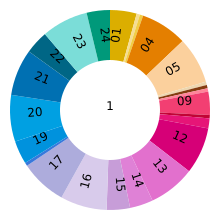}
            \caption{A+STU}
        \end{subfigure}
        \begin{subfigure}[b]{.45\linewidth}
            \includegraphics[width=\textwidth]{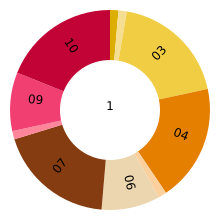}
            \caption{NA-NASR}
        \end{subfigure}
        \caption{Repartition of ballistic missile tracks}
        \label{repartition}
    \end{figure}

    MCDA is a mathematical tool that can be used to evaluate alternatives no a fixed number of metrics. To make these computations, each metric is normalized according to a utility function, that maps each possible value of the metric to $[0,1]$. The normalized values are then aggregated into higher level criteria, that are in turn aggregated. The result of this process is a tree whose root represents the overall performance of the system for each alternative, as represented on \cref{fig:elemttrack}. In standard MCDA methods, each metrics is only measured once for each alternative. In the case of the RRM though, there is an additional complexity, linked to the fact that the algorithm is run for multiple tracks, on multiple scenarios. As a consequence, there are many measures for the same metric. The tracks are evenly distributed across the scenarios, except for the ballistic missiles, whose distribution is provided on \cref{repartition}. Note that there are not the same number of scenarios for each RRM algorithm. For the nonadaptive algorithm (see \cref{Spb} for more information), only the relevant scearios are taken into account. In spite of applying the MCDA model to the averages, which has some drawbacks (see \cite{lab17v}), we propose a new approach, in which each value of each metruc is first normalized and then aggregated through an \emph{Ordered Weighted Average (OWA)} \cite{yag88,yag96}. This aggregation function can be used to put more or less weight on the worst evaluations, and therefore be more or less pessimistic or risk adverse
    
	More details on the MCDA method and elicitation approach used here can be found in \cite{labreuche2019}. The remaining of the paper is organized as follows: \cref{Spb} presents the naval scenario. Finally, \Cref{Sres} gives the results of the MCDA evaluation.
	
	
			\begin{sidewaysfigure*}[p]
		\centering
		\includegraphics[width=.95\textheight]{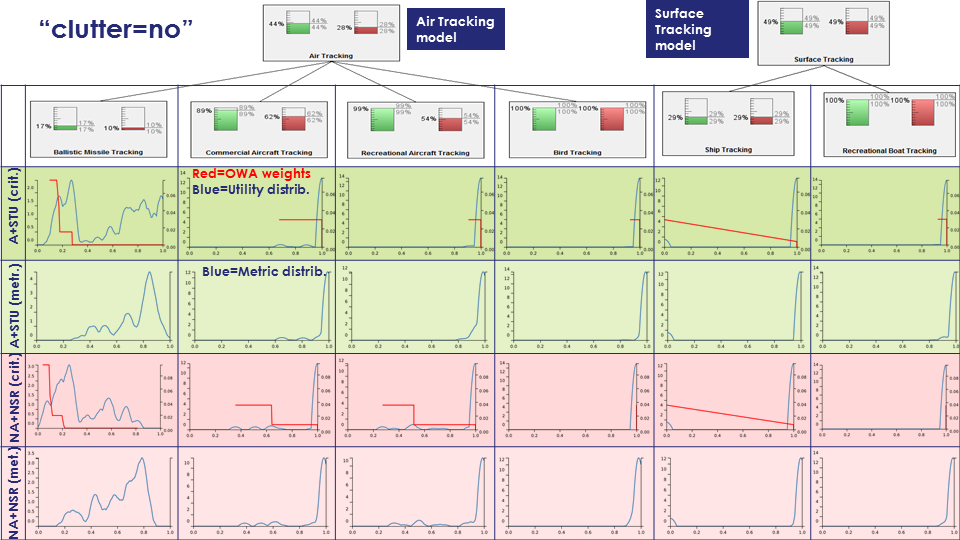}
		\caption{Details of the evaluations of track completeness without clutter.}
		\label{fig:opsc2}
	\end{sidewaysfigure*} 
	
	\begin{sidewaysfigure*}[!p]
		\centering
		\includegraphics[width=.95\textheight]{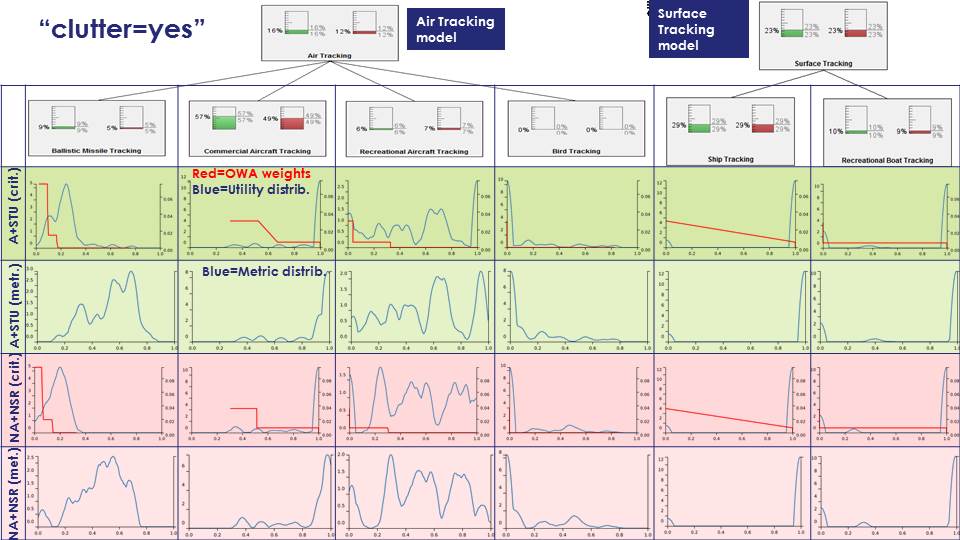}
		\caption{Details of the evaluations of track completeness in the presence of clutter.}
		\label{fig:opsc1}
	\end{sidewaysfigure*} 

	\section{Naval scenario}
	\label{Spb}
	
	\begin{figure*}[htbp]
	    \centering
	        \begin{subfigure}[b]{.32\linewidth}
	            \centering
	            \includegraphics[width=\linewidth]{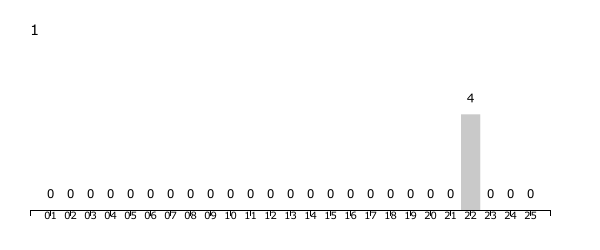}
	            \caption{Ballistic missiles}
	        \end{subfigure}
	        \begin{subfigure}[b]{.32\linewidth}
	            \centering
	            \includegraphics[width=\linewidth]{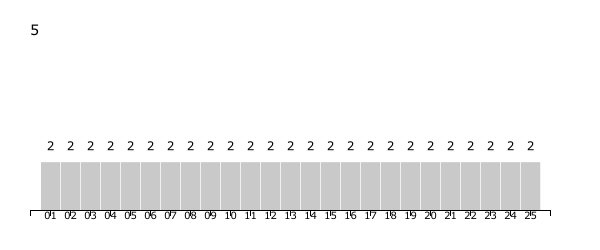}
	            \caption{Ships}
	        \end{subfigure}
	        \begin{subfigure}[b]{.32\linewidth}
	            \centering
	            \includegraphics[width=\linewidth]{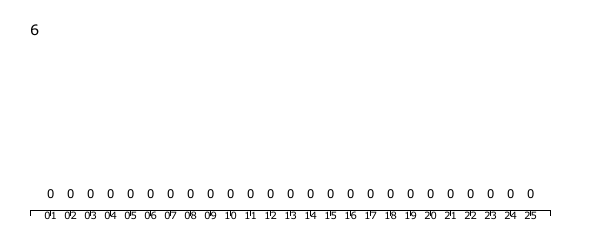}
	            \caption{Recreational boats}
	        \end{subfigure}\\
	        \begin{subfigure}[b]{.32\linewidth}
	            \centering
	            \includegraphics[width=\linewidth]{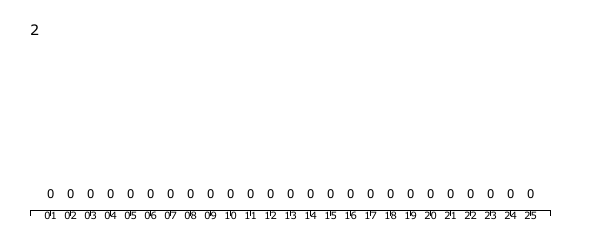}
	            \caption{Commercial aircrafts}
	        \end{subfigure}
	        \begin{subfigure}[b]{.32\linewidth}
	            \centering
	            \includegraphics[width=\linewidth]{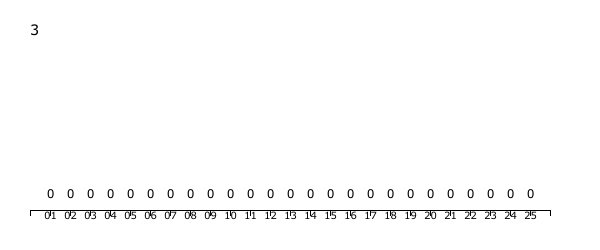}
	            \caption{Recreational aircrafts}
	        \end{subfigure}
	        \begin{subfigure}[b]{.32\linewidth}
	            \centering
	            \includegraphics[width=\linewidth]{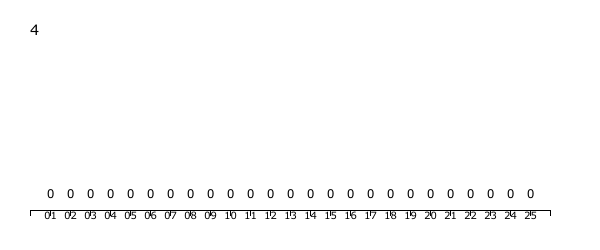}
	            \caption{Birds}
	        \end{subfigure}
	    \caption{Number of untracked targets -- A+STU without clutter}
	    \label{untracked}
	\end{figure*}
	
	\begin{figure*}[htbp]
	    \centering
	        \begin{subfigure}[b]{.32\linewidth}
	            \centering
	            \includegraphics[width=\linewidth]{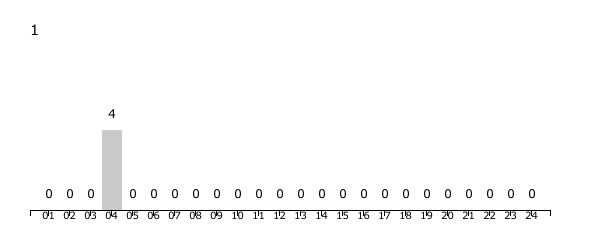}
	            \caption{Ballistic missiles}
	        \end{subfigure}
	        \begin{subfigure}[b]{.32\linewidth}
	            \centering
	            \includegraphics[width=\linewidth]{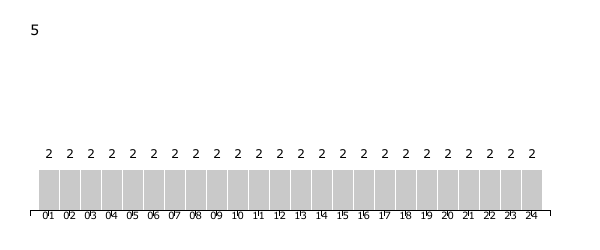}
	            \caption{Ships}
	        \end{subfigure}
	        \begin{subfigure}[b]{.32\linewidth}
	            \centering
	            \includegraphics[width=\linewidth]{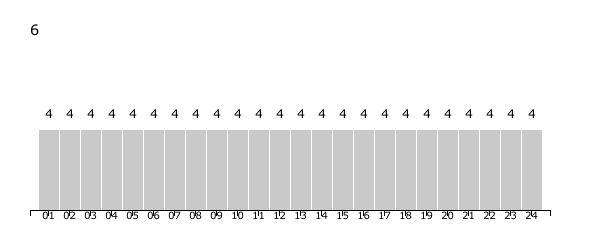}
	            \caption{Recreational boats}
	        \end{subfigure}\\
	        \begin{subfigure}[b]{.32\linewidth}
	            \centering
	            \includegraphics[width=\linewidth]{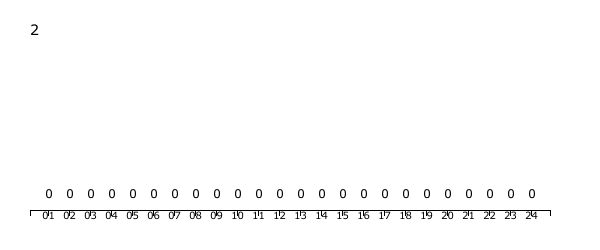}
	            \caption{Commercial aircrafts}
	        \end{subfigure}
	        \begin{subfigure}[b]{.32\linewidth}
	            \centering
	            \includegraphics[width=\linewidth]{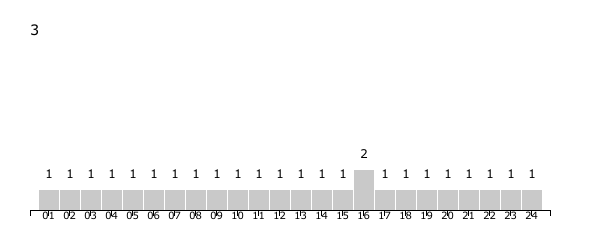}
	            \caption{Recreational aircrafts}
	        \end{subfigure}
	        \begin{subfigure}[b]{.32\linewidth}
	            \centering
	            \includegraphics[width=\linewidth]{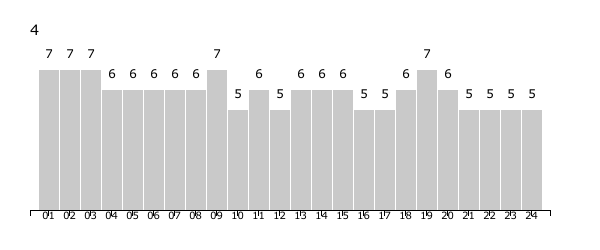}
	            \caption{Birds}
	        \end{subfigure}
	    \caption{Number of untracked targets -- A+STU with clutter}
	    \label{untracked-clutter}
	\end{figure*}
		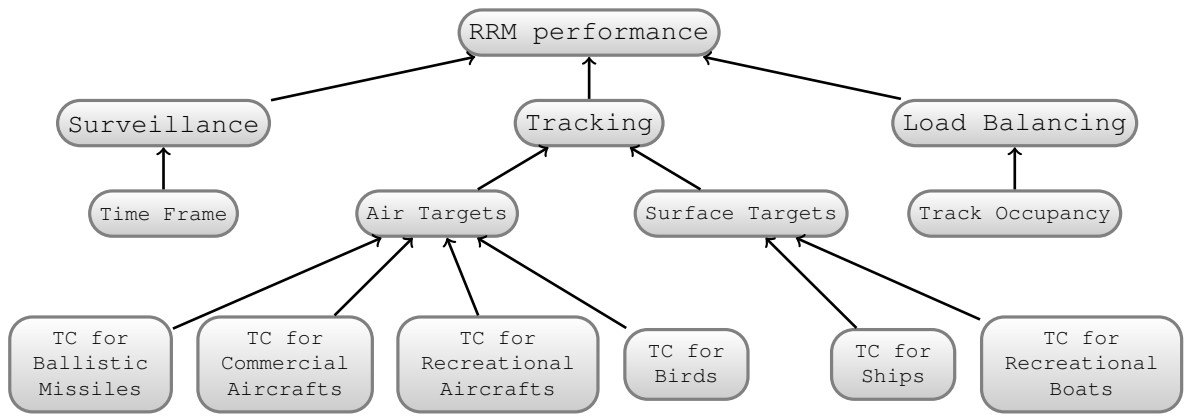
\begin{figure*}[h]
		
		\begin{center}
			
			\begin{tikzpicture}[scale=0.8,node distance=5mm,
			terminal/.style={
				rectangle,minimum size=6mm,rounded corners=3mm,
				very thick,draw=black!50,
				top color=white,bottom color=black!20,
				font=\ttfamily}]
			
			\tikzstyle{s}=[circle,draw, line width=1pt]
			
			\tikzstyle{s-out}=[circle,draw, double,  line width=1pt]
			
			\node (a4) at (4,1.5)  [terminal] {\small RRM performance};
			\node (b1) at (-3,0)  [terminal] {\small Surveillance};
			\node (b1b) at (-3,-1.5)  [terminal] {\scriptsize Time Frame};
			\node (b2) at (4,0)  [terminal] {\small Tracking};
			\node (b2a) at (1.5,-1.5)  [terminal] {\scriptsize Air Targets};
			\node (b2a1) at (-4.2,-4)  [terminal] {\scriptsize $\begin{array}{c} \mbox{TC for} \\ \mbox{Ballistic} \\ \mbox{Missiles} \end{array}$};
			\node (b2a2) at (-1,-4)  [terminal] {\scriptsize $\begin{array}{c} \mbox{TC for} \\ \mbox{Commercial} \\ \mbox{Aircrafts} \end{array}$};
			\node (b2a3) at (2.5,-4)  [terminal] {\scriptsize $\begin{array}{c} \mbox{TC for} \\ \mbox{Recreational} \\ \mbox{Aircrafts} \end{array}$};
			\node (b2a4) at (5.6,-4)  [terminal] {\scriptsize $\begin{array}{c} \mbox{TC for} \\ \mbox{Birds} \end{array}$};
			\node (b2c) at (6.5,-1.5)  [terminal] {\scriptsize Surface Targets};
			\node (b2c1) at (9,-4)  [terminal] {\scriptsize $\begin{array}{c} \mbox{TC for} \\ \mbox{Ships} \end{array}$};
			\node (b2c2) at (12.1,-4)  [terminal] {\scriptsize $\begin{array}{c} \mbox{TC for} \\ \mbox{Recreational} \\ \mbox{Boats} \end{array}$};
			\node (b3) at (11,0)  [terminal] {\small Load Balancing};
			\node (b3a) at (11,-1.5)  [terminal] {\scriptsize Track Occupancy};
			\draw [->,line width=1pt] (b1) -- (a4);
			\draw [->,line width=1pt] (b2) -- (a4);
			\draw [->,line width=1pt] (b3) -- (a4);
			\draw [->,line width=1pt] (b3a) -- (b3);
			\draw [->,line width=1pt] (b2a) -- (b2);
			\draw [->,line width=1pt] (b2a1) -- (b2a);
			\draw [->,line width=1pt] (b2a2) -- (b2a);
			\draw [->,line width=1pt] (b2a3) -- (b2a);
			\draw [->,line width=1pt] (b2a4) -- (b2a);
			\draw [->,line width=1pt] (b2c) -- (b2);
			\draw [->,line width=1pt] (b2c1) -- (b2c);
			\draw [->,line width=1pt] (b2c2) -- (b2c);
			\draw [->,line width=1pt] (b1b) -- (b1);
			\end{tikzpicture}
		\end{center}
		\vspace*{-0.2cm}
		\caption{Description of the elementary viewpoints on tracking performance, where TC stands for Track Completeness.}
		\label{fig:elemttrack}
	\end{figure*}
	
\begin{figure}[hbt]
		\centering
		\includegraphics[width=.45\textwidth]{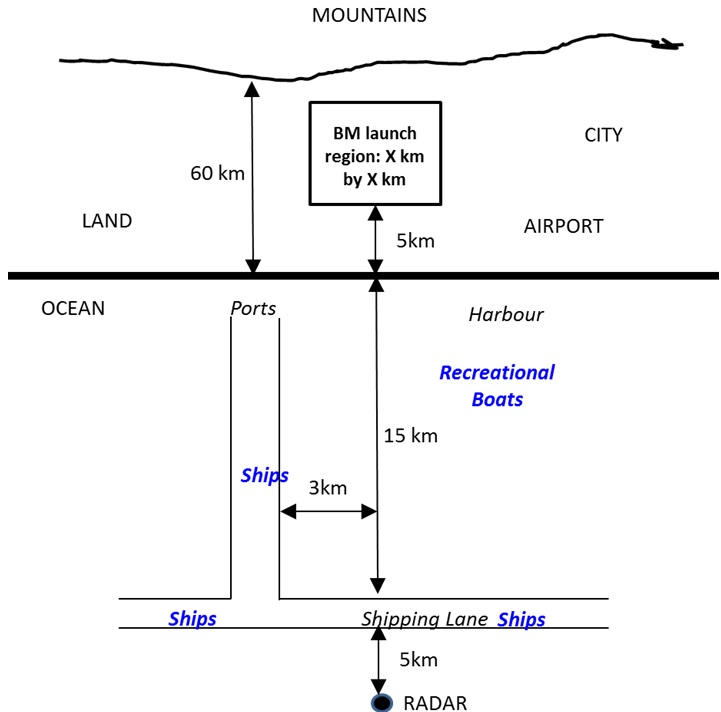}
		\caption{Top down view of a Ballistic Missile (BM) scenario}
		\label{fig:opsc}
	\end{figure}

	The problem at stake is to compare two solutions:
	\begin{itemize}
		\item \textbf{``NA+NSR''}: the standard legacy \emph{Non-Adaptive} algorithm, with No Special Rate;
		\item \textbf{``A+STU''}: the \emph{adaptive} algorithm with Special Track Update rates for the most threatening targets. 
		It includes three novel adaptive components: Fuzzy Logic Prioritization, Time Balancing Scheduling (TBS), and Adaptive Update Intervals for Tracking \cite{mooding15}.
	\end{itemize}
	The operational scenario is related to a Ballistic Missile Defence (BMD) mission, in which the ship contributes to the defence against a ballistic missile threat.  The radar’s role is to detect and track the ballistic missile.  
	The scenario is about radar tracking on a littoral region, with several clutter backgrounds including sea, land, and urban clutter -- see \Cref{fig:opsc}. In the scenario, targets are both surface and air targets:
	\begin{itemize}
	    \item surface targets include ships and recreational boats,
	    \item air targets consist of ballistic missiles, commercial aircrafts, recreational aircrafts, and birds.
	\end{itemize}
	
	The scenario is situated in a littoral region with a varied clutter background, including sea, land, and urban clutter -- see \Cref{fig:opsc}. 
	In the scenario, targets are both surface and air targets.
	Surface targets include ships and recreational boats.  
	Air targets consist of ballistic missiles, commercial aircrafts, recreational aircrafts, and birds.
	The ballistic missile launch region is shown as a square region over the land portion of the scenario. 
	As the area of the launch region increases, the radar will require more resources to detect and track a ballistic missile that is launched.
	The two solutions are evaluated against clutter and non-clutter conditions.
	Simulation has been performed with  Adapt\_MFR simulation tool.
	
	The evaluation is performed regarding:
	\begin{itemize}
		\item the ``Surveillance'' viewpoint, measured by metric ``Time Frame'', is the average refresh rate of surveillance over the whole space. The smaller the value of this metric, the better;
		\item the ``load balancing'' viewpoint, measured by metric ``Track Occupancy'', is the percentage of radar time used on tracking -- the complementary time being allocated to surveillance. The smaller the value of this metric, the better. 
		\item the ``Tracking'' viewpoint, measured by metric ``Track Completeness'', is the average percentage of each trajectory that is tracked. The larger the value of this metric, the better;
	\end{itemize}
	The expectations on track completeness and the severity of non-fulfillment on this metric depends on the type of targets.
	Hence, as proposed in \cite{lab17v}, we group together the values of track completeness for each type of target -- see \Cref{fig:elemttrack}.

	\section{Results}
	\label{Sres}
	
	Due to space limitations, we only present the details of results on tracking function. We present the results regarding the track completeness in \cref{fig:opsc2} for the case without clutter and in \cref{fig:opsc1} for the case with clutter.
	They are organized as an array. Columns represent the metrics: ``ballistic missile tracking'', ``commercial aircraft tracking'', ``recreational aircraft tracking'', ``bird tracking'', ``ship tracking'' and ``recreational boat tracking'' from left to right.
	The green (resp. red) represent the performance assessment between $0\%$ (criterion not satisfied at all) and $100\%$ (criterion perfectly satisfactory) for solution A+STU (resp. solution NA+STU).
	The two green (resp. red) rows represent the results for solution A+STU (resp. NA+NSR).
	The bottom green (resp. red) raw named ``A+STU (metr.) (resp. ``NA+NSR (metr.)'') shows an estimate of the probability distribution over the metric space. These are the raw values of metric ``track completeness'' from $0$ (worst value) to $1$ (best value) coming from the simulations.
	The top green (resp. red) raw named ``A+STU (crit.) (resp. ``NA+NSR (crit.)'') shows two curves: an estimate of the probability distribution over the the normalized values of the metric (\textsl{i.e.} the utility function applied to the metric), in blue and the corresponding values of the OWA weights, in red. The estimate of the probability distribution is got using Kernel Density Estimation \cite{silverman2018density} over the collected data.
	
	Looking at the probability distribution on the metrics, we note that the values are quite good. For instance, for A+STU on ballistic missiles, the main peak is around 0.85.
	After the application of the utility function, the probability distribution becomes much less optimistic.
	For instance, for A+STU on ballistic missiles, the main peak is around 0.3 with some kind of plateau between 0.7 and 1.
	The overall score presented in the gauge is the integral of the product of the blue and the red curves.
	Note that the red curve (OWA weights) are decreasing, \textsl{i.e.} more weights are put on the worst tracked targets.
	This explains the bad evaluations on the gauges.
	
	\paragraph{Case where there is no clutter} There is significant improvement on A+STU compared to NA+NSR, except for Surface tracking for which the results are similar. A+STU gets very good performances on ``Commercial Aircraft Tracking'', ``Recreational Aircraft Tracking'' and ``Bird Tracking''. 
	This is impressive, especially considering the pessimistic OWA weights. On the other hand, there is no major improvement on ballistic missiles tracking. The values of this metric are relative wide spread on the $[0,1]$ range; hence there is not at least $90\%$ of well-tracked ballistic missiles, which explains the bad evaluation. Overall, A+STU has better performance in all 3 of the main criteria: tracking, surveillance and load balancing. Note that less tracking beams are needed to track targets for the adaptive RMM, which improves load balancing. This extra saved times allows it to improve the surveillance performance (time frame) over the non-adaptive RMM.
	
	\paragraph{Case where there is some clutter} The general comments are similar to the case without clutter. We see that the presence of clutter has very little consequence on surveillance and load balancing, which is a positive point. However, the tracking performances are much lower, especially for ``Recreational Aircraft Tracking'' and ``Bird Tracking''. Only ``Commercial Aircraft Tracking'' maintains a good performance (which is natural as these tracks are easier to track).
	
	The results regarding the number of untracked targets are presented in \cref{untracked,untracked-clutter}. Due to space limitation, only the results of the A+STU RRM algorithm is presented. The difference between the case with and without clutter is identical for the A+STU RRM algorithm. It is worth noting that there are some differences with the track completeness. While there are some degradation regarding track completeness for ballistic missiles between the case without/with clutter, the number of untracked missiles stays the same. The same goes for the number of untracked ships and commercial aircrafts. The changes regarding the other targets are consistent with the previous results. These elements are quite interesting: they indicate that, though the completeness of the tracks deteriorate, some targets keep being tracked by the radars, in particular the ballistic missiles and the ships. Note however that, though the figures are not represented here, it is also the case for the NA+NSR alternative.
	
	\section{Conclusion}
	\label{Scl}
	
	By integrating preferences of the decision maker(s) on their expectations and priorities, our MCDA approach helps connecting the experts' operational perspective on solutions to end users.
	Furthermore, it provides transparency on the evaluation process and provides explainations of the figures to the user, which is important to get acceptance and trust from them.
	
	Future works may include the extension of our model towards tests of actual radars. In this case, we will need to handle statistical values regarding the figures of merits; in such situation, a more stable RRM algorithm (resulting in a distribution with less variability), may be preferred. We could integrate an additional metric regarding the dispersion of the figures to the MCDA model. It would also be possible to extend our MCDA tool in order to propagate distributions rather than raw scores.
	
	\bibliographystyle{plain}
	\bibliography{biblio}
	

\end{document}